\documentclass[12pt,epsf,amssymb]{article}
\usepackage{tabularx}
\usepackage{array}
\usepackage{graphics}
\usepackage{epsfig}

\makeatletter

\usepackage{verbatim}

\newcommand{\msbar}{\overline{\rm MS}}
\newcommand{\bea}{\begin{eqnarray}}
\newcommand{\eea}{\end{eqnarray}}
\newcommand{\beq}{\begin{equation}}
\newcommand{\eeq}{\end{equation}}
\newcommand{\gev}{{\rm GeV}}

\newcommand{\pdir}{p\kern -5.2pt\raise 0.2ex\hbox {/}}
\newcommand{\vdir}{v\kern -5.75pt\raise 0.15ex\hbox {/}}
\newcommand{\kdir}{k\kern -5.75pt\raise 0.15ex\hbox {/}}
\newcommand{\epsdir}{\epsilon\kern -5.0pt\raise 0.15ex\hbox {/}}
\newcommand{\bvdir}{\bar{v}\kern -5.75pt\raise 0.15ex\hbox {/}}
\newcommand{\Ddir}{D\kern -7.75pt\raise 0.20ex\hbox {/}}
\newcommand{\ldir}{l\kern -5.0pt\raise 0.2ex\hbox{/}}
\newcommand{\varepsdir}{\varepsilon\kern -5.5pt\raise 0.15ex\hbox{/}}

\newcommand{\kkbar}{K^0-{\overline{K^0}}}
\makeatother

\begin{document}
\thispagestyle{empty}
\begin{flushright}
\begin{tabular}{l}
ROMA-1306/00\\
\end{tabular}
\end{flushright}
\vskip 3.0cm\par
\begin{center}
{\par\centering \textbf{\LARGE An Estimate of }}\\
\vskip 0.2cm\par
{\par\centering \textbf{\LARGE the $\kkbar$ Mixing Amplitude}}\\
\vskip 1.05cm\par
{\par\centering \large
\sc  Damir~Becirevic, Davide~Meloni, Alessandra~Retico}
{\par\centering \vskip 0.9 cm\par}
{\sl
 Dipartimento di Fisica, Universit\`a di Roma ``La Sapienza" \\
and INFN, Sezione di Roma,\\
Piazzale Aldo Moro 2, I-00185 Rome, Italy. \\
\vspace{.25cm}
}
\vskip1.cm                                                                      
{\vskip 0.25cm\par}
\end{center}
 
\vskip 0.55cm
\begin{abstract}
We computed the $B_K$ parameter on the lattice by using a non-perturbatively 
improved Wilson action.
From our quenched simulation we obtain 
$B_K^{\msbar}(2 \gev)=0.73(7)^{+0.05}_{-0.01}$.
To gain some insight in the systematic errors due to the use of the 
quenched approximation,
we also made the unquenched computation with Wilson fermions. 
We do not observe 
any significant deviation with respect to the quenched result.
\end{abstract}
\setcounter{page}{1}                                                            
\noindent
 
\renewcommand{\thefootnote}{\arabic{footnote}}
 
\newpage
\setcounter{page}{1}
\setcounter{footnote}{0}
\setcounter{equation}{0}
\section{Preliminaries: Motivation and Status}
The experimental value of the parameter $\varepsilon_K$, 
which is the measure of indirect CP violation in the kaon system, has been accurately determined~\cite{Groom:2000in}.
After combining $\vert \varepsilon_K^{\rm (exp.)}\vert $ with its theoretical expression
derived in the Standard Model (SM),
one obtains a hyperbola in the ($\bar \rho$, $\bar \eta$) 
plane~\cite{Buras:1998ra,Buchalla:1996vs},
\bea \label{UNIT}
\bar \eta A^2\ \hat B_K\ \biggl( \, 1.11(5) \cdot A^2 ( 1 - \bar \rho)\ +\ 0.31(5)\, \biggr) =
0.226\;,
\eea
which, together with two equations involving $B$-mesons, represents one of the   
main constraints onto the shape of the CKM unitarity
triangle. In eq.~(\ref{UNIT}), the quantities $\lambda =0.22$, $A$, $\bar \eta$ , $\bar \rho$ are the 
standard Wolfenstein parameters of the CKM matrix elements~\cite{Wolfenstein:1983yz,Buras:1994ec}, 
while $\hat B_K$ is the so-called bag parameter which encodes the low 
energy QCD contribution to the $\kkbar$ mixing amplitude. The main uncertainty 
in eq.~(\ref{UNIT}) comes precisely from $\hat B_K$ (to which we focus in this
letter) and to less 
extent to the parameter $A$ which is known from $\vert V_{cb}\vert = A \lambda^2$.

We first remind the reader the definition of $\hat B_K$.
As introduced in ref.~\cite{Gaillard:1974nj}, the parameter 
$\hat B_K$ measures a mismatch between the hadronic matrix 
element responsible for the $\kkbar$ mixing and its value obtained
by applying the vacuum saturation approximation (VSA), {\it i.e.}
\bea
\label{def1}
\langle \bar K^0 \vert Q(\mu) \vert K^0 \rangle = {8\over 3} f_K^2 m_K^2 B_K(\mu)\  ,
\eea
where $\mu$ indicates the scale dependence of the operator 
$Q=\bar s \gamma_\mu (1 - \gamma_5) d\ \bar s \gamma_\mu (1 - \gamma_5) d$, in 
some renormalization scheme. The QCD anomalous dimension of this operator, 
in the RI-MOM and $\msbar$ schemes, is known 
from perturbation theory to NLO accuracy~\cite{Ciuchini:1998bw}--\cite{Buras:1990fn}. To that 
precision, $B_K(\mu)$ can be converted to the renormalization group 
invariant (RGI) form,
\bea
\label{anomalous}
\hat  B_K     = 
\alpha_s(\mu)^{-\gamma_0/2\beta_0} 
\left( 1 + {\alpha_s(\mu) \over 4 \pi} J\right) \ B_K(\mu) 
\ , 
\eea
which is the one  that is needed in eq.~(\ref{UNIT}). 
Besides the scheme independent $\gamma_0 = 4$ and $\beta_0 = 11 - 2 n_f/3$, ($n_f$ is the number of active
flavours), the NLO coefficient $J$ is renormalization scheme dependent and in
the two 
schemes which we use in this letter, it is given by~\cite{Ciuchini:1998bw}--\cite{Buras:1990fn} 
\bea
&& J_{\rm RI-MOM} \ = \ 8 \log 2\ -\ {\; 17 397 -2070 n_f + 104 n_f^2\; 
 \over 6 ( 33 - 2 n_f)^2}\  \,  ,\cr
&& \cr
&& J_{\msbar} \ = \ {\; 13095 - 1626\,n_f + 8\,n_f^2\;  \over 6\, \left( 33 - 2\,n_f \right)^2} \, \, .
\eea
The main task is to compute $B_K(\mu)$ in some scheme, by using a non-perturbative 
technique, and then to convert the result to the RGI form, $\hat B_K$. We choose
 lattice QCD and the RI-MOM scheme.

Before discussing our computation, let us briefly recapitulate the actual situation 
concerning the value of $B_K$, which at the same time gives a good motivation
to compute this quantity.

Historically, the three distinct approximations applied to this problem lead 
to different results: VSA gave $B_K = 1$~\cite{Gaillard:1974nj},  
PCAC and SU(3) symmetry implied  $\hat B_K = 1/3$~\cite{Donoghue:1982cq}, and
the limit of the large number of colours ($N_c\to \infty$) yielded 
$B_K = 3/4$~\cite{Bardeen:1988vg}.~\footnote{Note that VSA and large $N_c$ cannot
distinguish between $B_K(\mu)$ and $\hat B_K$.}
Such a large spread of values prompted many physicists to develop various 
models which typically combine the large $N_c$ expansion with chiral
perturbation theory ($\chi$PT), plus some additional assumption(s).
Two most recent studies, in which such a strategy has been adopted, resulted in  
$\hat B_K = 0.41(9)$~\cite{Peris:2000sw} and 
$\hat B_K = 0.77(5)(5)$~\cite{Bijnens:2000zn}~\footnote{
The result of ref.~\cite{Peris:2000sw} is obtained in the chiral limit in which
the authors of ref.~\cite{Bijnens:2000zn} quote 
 $\hat B_K^\chi = 0.32(6)(12)$.}. 
On the other hand, the chiral quark model prediction of ref.~\cite{Bertolini:1998ir} favours much 
larger value, {\it i.e.} $\hat B_K = 1.1(2)$. 
QCD sum rules have also been used to compute $\hat B_K$, with the results
dispersed over a large range of values.~\footnote{Three different QCD sum rule techniques
lead to: $\hat B_K = 0.5(1)(2)$~\cite{Bilic:1988gk},  
$0.84(8)$~\cite{Reinders:1987gv},  $0.39(10)$~\cite{Prades:1991sa}.}

Clearly, these refinements of the approximate results did 
not help clarifying the issue of the correct value for $\hat B_K$. 
It should be mentioned, however, that in SM one can extract the range 
of allowed values for $\hat B_K$, by requiring the consistency 
with the $B$-physics constraints 
on the CKM triangle. In this way, the authors of ref.~\cite{Ciuchini:2000de} 
found that $0.62 \leq \hat B_K \leq 1.48$ (with $95\%$ C.L.), thus excluding 
the low SM values such as $\hat B_K < 0.6$ (see \cite{Ciuchini:2000de}
and references therein).

The most direct and systematically most improvable way to compute 
the matrix element~(\ref{def1}) in QCD is by using the lattice simulations. 
Over the past decade, a substantial progress in controlling various systematic
uncertainties of this approach has been made (see refs.~\cite{Sharpe:1997ih,Lellouch:2000bm,Lubicz:2000dj} for
reviews which contain a complete list of references).

Three currently used formulations of the lattice fermion action have been 
employed 
in the computation of $\hat B_K$ (or, as it is customary 
$B_K^{\msbar}(2\ \gev)$) , and in the quenched approximation, they 
lead to the following results:
\begin{itemize}
\item With the staggered fermions one can compute the matrix 
element~(\ref{def1}) by using 
very light quark masses. The renormalization of the operator $Q(\mu)$ has 
been performed perturbatively. By extrapolating to the continuum limit, 
JLQCD~\cite{Aoki:1998nr} obtained
\bea
B_K^{\msbar}(2\ \gev)=0.628(42)\ ,\eea 
and thus improved (confirmed) the previous result of ref.~\cite{Kilcup:1998ye}, 
$B_K^{\msbar}(2\ \gev)=0.62(2)(2)$.
\item With Wilson fermions, as compared to the staggered, one cannot work with the very light 
quark masses, but the renormalization of the operator $Q$ (which is more 
delicate due to the
mixing with the operators of the same dimension) has been done non-perturbatively. 
Presently, the most refined result is the one by JLQCD~\cite{Aoki:1999gw},
\bea
B_K^{\msbar}(2\ \gev)=0.69(7)\ ,\eea 
fully consistent with the previous 
$B_K^{\msbar}(2\ \gev)=0.66(11)$, as obtained in ref.~\cite{Conti:1998qk}.
\item
Very recently, also the domain wall fermions were employed in this
computation. 
The results of RBC and CP-PACS are:
\bea
&&B_K^{\msbar}(2\ \gev)=0.538(8)~\cite{Blum:2000tt}\,, \nonumber \\ 
&&B_K^{\msbar}(2\ \gev)=0.575(6)~\cite{AliKhan:2000yb}\,.
\eea 
Notice that RBC also implemented the non-perturbative renormalization in their 
computation~\cite{Dawson:2000kh,Zhestkov:2000bs}.
\end{itemize}
In summary, from the use of various lattice techniques, one concludes that the value of 
$B_K^{\msbar}(2\ \gev)$ lies in the interval $0.53 \leq B_K^{\msbar}(2\ \gev) \leq 0.77$, or 
in the RGI form, $0.72 \leq \hat B_K \leq 1.05$.~\footnote{This range of values
agrees with the one presented by L.~Lellouch at ``Lattice
2000'', namely $0.66 \leq \hat B_K \leq 1.06$~\cite{Lellouch:2000bm}.}

In this letter, we contribute the ongoing discussion 
by presenting the results of our computation of
the matrix element~(\ref{def1}), for which we use the ${\cal O}(a)$ 
non-perturbatively improved Wilson action~\cite{Luscher:1997ug}, and 
renormalize the operator $Q(\mu)$ non-perturbatively. The operator 
$Q(\mu)$ has not been improved, though.~\footnote{As compared to 
ref.~\cite{Conti:1998qk}, our action is non-perturbatively improved 
whereas their was tree-level improved. In addition, we do not rotate
the quark fields. This rotation is unnecessary (the 4-fermion operator 
is not improved) and it introduces a statistical noise. This is why our
statistical errors are smaller than theirs.}
In order to examine the systematic uncertainty arising from the 
use of quenched
approximation, we attempted an unquenched computation
with two degenerate dynamical quarks ($n_f=2$) and for two choices of the
dynamical quark masses. From the direct comparison
to the quenched result, and bearing in mind the modest statistical 
quality of our unquenched data, we do not see any difference between 
the quenched and unquenched results.

Our final numbers are:
\bea
B_K^{\msbar}(2\ \gev) = 0.73(7)^{+0.05}_{-0.01};\quad \hat B_K=1.01(9)^{+0.07}_{-0.01}\; ,
\eea
thus favouring the larger values for this quantity.

\vspace*{6mm}

\section{Outline of the Lattice Computations}
\label{sec:setup}
In order to determine the value of the parameter $B_K(\mu)$ in the RI-MOM scheme, we studied 
the relevant matrix element on four lattices for which the main features are summarized in tab.~\ref{tab1}. 
For technical details related to the production of the
unquenched configurations, consult refs.~\cite{Lippert:1998qy,Eicker:1999sy}. 
The values of the Wilson hopping parameters are listed in tab.~\ref{tab2}, 
together with the corresponding 
masses of pseudoscalar mesons and ratios of the pseudoscalar to vector meson masses. 
In our notation the meson masses in lattice units are written in capital letters, 
whereas the small case letters are used to denote the same masses in physical units 
({\it e.g.} $M_P = a m_P$). As usual, 
the meson masses are extracted from the exponential fall-off of the two-point 
correlation functions. 
All the source operators used in this work are local. Statistical errors are estimated 
by using the jackknife procedure.
For completeness,  we give for each lattice the value of the critical hopping parameter
as obtained
by linearly extrapolating the square of the pseudoscalar meson mass ({\it i.e.} the 
valence quark mass) to zero. Using the same labels as in tab.\ref{tab1}, we give:
\bea
\vspace*{-6mm}
& & \quad \quad {\scriptstyle \rm (a)}  \hspace*{1.7cm} {\scriptstyle \rm (b1)} \hspace*{1.75cm} {\scriptstyle \rm (b2)} \hspace*{1.7cm} 
 {\scriptstyle \rm (c)}\cr
\kappa_{crit} &=& \left\{ 0.13580(2),\ 0.15931(6),\ 0.15885(7),\ 0.15494(5)  \right\} \; ,
\eea

The {\sl physical} volume of the lattice (c) 
has been chosen to be nearly equal to the one used in simulations with partial 
unquenching ($L^3\approx (1.9\ {\rm fm})^3$). Moreover, the hopping 
parameters 
for run (c) are chosen in such a way that the masses of pseudoscalar mesons
are very close to the ones obtained from the unquenched runs. In this way any sizable 
difference between the physical quantities obtained from the quenched and 
unquenched simulations would signal the quenching error.

\begin{table} 
\begin{center} 
\begin{tabular}{|c|c|c|c|c|c|}
\hline
{\phantom{\Huge{l}}}\raisebox{-.2cm}{\phantom{\Huge{j}}}
{\sf Label}&$\left(\beta ,\ L^3\times T \right)$& $\mathsf c_{\mathsf{SW}}$  &  {\sf Statistics} &
$a^{-1}(m_{K^\ast})[\gev ]$ &   {\sf Comment} \\
\hline 
{\phantom{\Huge{l}}}\raisebox{-.1cm}{\phantom{\Huge{j}}}
(a)&$\left( 6.2,\ 24^3\times 48\right)$&  $1.614$  &  200 config.'s& 2.71(8)   &Quenched\\
{\phantom{\Huge{l}}}\raisebox{-.1cm}{\phantom{\Huge{j}}}
(b1)&$\left( 5.6,\ 24^3\times 40\right)$& 0 &  60 config.'s& 2.48(11)  
& P.U. ($\kappa_{sea}\! = 0.1575$) \\
{\phantom{\Huge{l}}}\raisebox{-.1cm}{\phantom{\Huge{j}}}
(b2)&$\left( 5.6,\ 24^3\times 40\right)$&  0&  60 config.'s & 2.60(12)  
& P.U. ($\kappa_{sea}\! = 0.1580$) \\
{\phantom{\Huge{l}}}\raisebox{-.1cm}{\phantom{\Huge{j}}}
(c)&$\left( 6.1,\ 24^3\times 40\right)$& 0&  60 config.'s   & 2.51(7)
& Quenched 
 \\ \hline
\end{tabular}
\caption{\label{tab1}
\small{\sl Characteristics of the lattices used in this work. P.U. stands for the partial
unquenching where the value of the corresponding dynamical (sea) quark mass parameter is indicated
in the parentheses. $a^{-1}$ has been estimated from the mass of the $K^\ast$ meson in a way 
explained in ref.~{\rm \cite{Becirevic:1998jp}}.}}
\end{center}
\vspace*{-.3cm}
\end{table}
\begin{table}[h!]
\begin{center} 
\begin{tabular}{|c|c|c|c|c|}
\cline{2-5}
\multicolumn{1}{l|}{}&{\phantom{\Huge{l}}}\raisebox{-.1cm}{\phantom{\Huge{j}}}\underline{$\kappa_1$}& \underline{$\kappa_2$}& \underline{$\kappa_3$}& \underline{$ \kappa_4 $}  \\
\hline 
{\phantom{\Huge{l}}}\raisebox{-.1cm}{\phantom{\Huge{j}}}
\underline{\sf{(a)}}& \sf 0.1352 & \sf 0.1349 & \sf 0.1344 & -- \\
{\phantom{\Huge{l}}}\raisebox{-.1cm}{\phantom{\Huge{j}}}
$M_P$ & 0.200(2) & 0.245(2) & 0.307(1) & -- \\  
{\phantom{\Huge{l}}}\raisebox{-.1cm}{\phantom{\Huge{j}}}
$m_P/m_V$ & 0.592(23) & 0.678(16) & 0.758(9) & -- \\ \hline 
{\phantom{\Huge{l}}}\raisebox{-.1cm}{\phantom{\Huge{j}}}
\underline{\sf{(b1)}}& \sf 0.1580 & \sf 0.1575 & \sf 0.1570 & \sf 0.1560 \\
{\phantom{\Huge{l}}}\raisebox{-.1cm}{\phantom{\Huge{j}}}
$M_P$ & 0.230(5) & 0.268(5) & 0.303(4) & 0.366(4) \\  
{\phantom{\Huge{l}}}\raisebox{-.1cm}{\phantom{\Huge{j}}}
$m_P/m_V$ & 0.612(30) & 0.685(22) & 0.735(17) & 0.802(11) \\ \hline 
{\phantom{\Huge{l}}}\raisebox{-.1cm}{\phantom{\Huge{j}}}
\underline{\sf{(b2)}}& \sf 0.1580 & \sf 0.1575 & \sf 0.1570 & \sf 0.1560 \\
{\phantom{\Huge{l}}}\raisebox{-.1cm}{\phantom{\Huge{j}}}
$M_P$ & 0.190(7) & 0.234(5) & 0.273(4) & 0.342(3) \\  
{\phantom{\Huge{l}}}\raisebox{-.1cm}{\phantom{\Huge{j}}}
$m_P/m_V$ & 0.633(15) & 0.700(16) & 0.748(15) & 0.816(11) \\ \hline 
{\phantom{\Huge{l}}}\raisebox{-.1cm}{\phantom{\Huge{j}}}
\underline{\sf{(c)}}& \sf 0.15363 & \sf 0.15338 & \sf 0.15311 & \sf 0.15258 \\
{\phantom{\Huge{l}}}\raisebox{-.1cm}{\phantom{\Huge{j}}}
$M_P$ & 0.228(5) & 0.248(4) & 0.269(4) & 0.307(4) \\  
{\phantom{\Huge{l}}}\raisebox{-.1cm}{\phantom{\Huge{j}}}
$m_P/m_V$ & 0.623(12) & 0.662(11) & 0.698(10) & 0.752(8) \\ \hline 
\end{tabular}
\caption{\label{tab2}
\small{\sl Pseudoscalar meson masses (in lattice units), and the ratio of the
pseudoscalar-to-vector mesons consisting of two degenerate (valence) quarks
of mass corresponding to the indicated value of the hopping parameter $\kappa_i$. 
Labels (a,b1,b2,c) refer to the lattices with characteristics given in 
tab.~{\rm \ref{tab1}}.}}
\end{center}
\vspace*{-.3cm}
\end{table}
 
Let us now turn to the definition of the operator $Q(\mu)$ that 
appears in eq.~(\ref{def1}).
Since the Wilson term in the fermion action explicitly breaks the chiral 
symmetry, an extra mixing of the operator $Q$ with other dimension-6 operators
 occurs. Consequently, in order to match the bare operator 
computed on the lattice to the continuum one, one has to subtract the 
effects of this mixing. In addition, one also needs the overall multiplicative 
renormalization constant. In a compact form, this procedure can be written as~\cite{Crisafulli:1996ad,Donini:1999sf} 
\bea
\label{subtr}
Q(\mu) = Z (\mu, g_0^2)
\ \left( Q_1 + \sum_{i\neq 1} \Delta_i(g_0^2)  Q_i \right)\ ,
\eea
where $Q_i$ on the r.h.s. denotes a bare lattice operator from the basis of 4-fermion operators, 
which we choose to be:
\bea 
\label{basis}
 Q\equiv Q_1 &=& \bar s \gamma_\mu  d \,
\bar s \gamma_\mu  d + \bar s \gamma_\mu  \gamma_{5}  d \,
\bar s \gamma_\mu \gamma_{5}  d \nonumber \\
{\phantom{\Huge{l}}}\raisebox{-.05cm}{\phantom{\Huge{j}}}  Q_2 &=& \bar s \gamma_\mu  d \,
\bar s \gamma_\mu  d - \bar s \gamma_\mu  \gamma_{5}  d \,
\bar s \gamma_\mu \gamma_{5}  d  \nonumber\\
{\phantom{\Huge{l}}}\raisebox{-.05cm}{\phantom{\Huge{j}}}  Q_3 &=& \bar s   d \,
\bar s  d - \bar s \gamma_{5}  d \,
\bar s  \gamma_{5}  d   \\
{\phantom{\Huge{l}}}\raisebox{-.05cm}{\phantom{\Huge{j}}}  Q_4 &=& \bar s  d \,
\bar s d + \bar s   \gamma_{5}  d \,
\bar s  \gamma_{5}  d   \nonumber\\
{\phantom{\Huge{l}}}\raisebox{-.05cm}{\phantom{\Huge{j}}} Q_5 &=& \bar s \sigma_{\mu\nu}  d \,
\bar s \sigma_{\mu\nu}  d  \;. \nonumber 
\eea 
To compute the subtraction  ($\Delta_i(g_0^2)$) and the
renormalization ($Z (\mu, g_0^2)$) constants non-perturbatively,   
we use the technique proposed and elaborated in refs.~\cite{Crisafulli:1996ad,Donini:1999sf}, 
and work in the RI-MOM scheme in the Landau gauge. 
In tab.~\ref{tab3}, we list the non-perturbatively determined values for 
$Z (\mu, g_0^2)$ and $\Delta_i(g_0^2)$
for all four lattices. The actual computation is performed at two values of the renormalization 
scale, $\mu a = 0.7, 1.4$, and for each value of the quark mass. For the presentation convenience, 
we linearly extrapolate each constant to the chiral limit, $\kappa_q \to \kappa_{crit}$. This 
extrapolation is very smooth. 
In addition, we evolve $Z(\mu a)$ to the scale $\mu a=1$ 
by using the NLO anomalous dimension coefficient (see eq.~(\ref{anomalous})) 
and by taking the 
two-loop expression
for $\alpha_s(\mu)$, normalized at $\alpha_s(3.41/a)$ by the value obtained from
the average plaquette, as prescribed in 
refs.~\cite{Lepage:1993xa,Davies:1997mg}.~\footnote{We stress that the values of $Z(\mu)$ which are
computed at two values of the renormalization scale, are consistent with the NLO anomalous dimension~(\ref{anomalous}). 
The difference that arise from the evolution from the two scales to $\mu=1/a$ is smaller than the
statistical error and it has been included in the errors quoted in tab.~\ref{tab3}.} 
\begin{table}[t!] 
\begin{center} 
\begin{tabular}{|c|c|c|c|c|c|c|} 
\hline
{\phantom{\Huge{l}}}\raisebox{-.2cm}{\phantom{\Huge{j}}}
{\sl Label}& {\sl Method }  & $Z(\mu=1/a , g_0^2)$   & $\Delta_2(g_0^2)$ & $\Delta_3(g_0^2)$ & $\Delta_4(g_0^2)$ & $\Delta_5(g_0^2)$ \\   \hline \hline 
{\phantom{\Huge{l}}}\raisebox{-.2cm}{\phantom{\Huge{j}}}
\underline{\sf{(a)}} &NP &  
 0.62(1) & -0.058(6)& -0.018(3)& 0.016(2) &0.003(2) \\ 
{\phantom{\Huge{l}}}\raisebox{-.2cm}{\phantom{\Huge{j}}}
 &PT &  
 0.68        & -0.041& -0.008& 0.015 &0.015   \\ \hline
{\phantom{\Huge{l}}}\raisebox{-.2cm}{\phantom{\Huge{j}}}
\underline{\sf{(b1)}} &NP &  
0.51(2) & -0.198(16)& -0.030(8)& 0.060(17) &0.019(11)    \\ 
{\phantom{\Huge{l}}}\raisebox{-.2cm}{\phantom{\Huge{j}}}
 &PT &  
 0.45        & -0.231& -0.042& 0.084 &0.084     \\ \hline
{\phantom{\Huge{l}}}\raisebox{-.2cm}{\phantom{\Huge{j}}}
\underline{\sf{(b2)}} &NP &  
0.52(2) & -0.203(12)& -0.039(8)& 0.066(18) &0.013(10)    \\ 
{\phantom{\Huge{l}}}\raisebox{-.2cm}{\phantom{\Huge{j}}}
 &PT &  
 0.45        & -0.230& -0.042& 0.083 &0.083     \\ \hline
{\phantom{\Huge{l}}}\raisebox{-.2cm}{\phantom{\Huge{j}}}
\underline{\sf{(c)}} &NP &  
0.55(2)  & -0.152(8)& -0.028(6)& 0.047(10) &0.007(5)     \\ 
{\phantom{\Huge{l}}}\raisebox{-.2cm}{\phantom{\Huge{j}}}
 &PT &  
 0.52        & -0.173& -0.032& 0.063 &0.063     \\ \hline
\end{tabular} 
\caption{\label{tab3}
\small{\sl Renormalization and subtraction constants (eq.~(\ref{subtr})),
evaluated non-perturbatively in the (Landau) RI-MOM scheme,  extrapolated 
(linearly in $m_q$) to the chiral limit and evolved to $\mu = 1/a$
by using the NLO evolution. Non-perturbative results are 
directly confronted to the perturbative values. The labels coincide 
with those used in tab.~\ref{tab1}.}}
\end{center}
\vspace*{-.3cm}
\end{table}
At the scale $\mu = 1/a$
we make a comparison with the results of one-loop perturbation theory. 
We use the perturbative expressions of ref.~\cite{Gupta:1997yt}, and combine them with the results of
ref.~\cite{Capitani:1997xj}. In the basis~(\ref{basis}) and in
the RI-MOM scheme they read:
\bea
&&Z(1/a, g_0^2) = 1 - {g_0^2\over 4 \pi} ( 3.674 - 0.742 c_{SW} - 0.388 c_{SW}^2)\ ,\cr
&& \cr 
&&Z(1/a, g_0^2)\cdot \Delta_i(g_0^2) = c_i {g_0^2\over 4 \pi} ( 0.767 - 0.795 c_{SW} + 0.272 c_{SW}^2)\ ,
\eea
where $c_2=-11/12$, $c_3=-1/6$ , $c_4=c_5=1/3$.
In the numerical computations, we use the boosted coupling, 
$g_0^2\to g_0^2=6/(\beta \langle P\rangle)$~\footnote{In the improved case, we
use $c_{SW}=1$, to be consistent with the 1-loop order perturbation theory.}, where the average plaquette values are computed in our simulations:
\bea
& & \quad \; {\scriptstyle  \rm (a)}  \hspace*{1.1cm} {\scriptstyle \rm (b1)} 
\hspace*{1.1cm} {\scriptstyle \rm (b2)} \hspace*{0.7cm} 
 {\scriptstyle \rm (c)}\cr
\langle P\rangle &=& \left\{ 0.6136,\; 0.5725,\; 0.5735,\; 0.6050  \right\}
\quad \,.
\eea
The labels above figures are the same as in tab.~\ref{tab1}.
From tab.~\ref{tab3} we see that the non-perturbatively computed subtraction 
constants $\Delta_i$ are not very far from their perturbatively estimated 
values (except for the case of $\Delta_5(g_0^2)$ whose perturbative value is
generally much larger than the non-perturbative one).
The value of the non-perturbatively computed renormalization constant  
differs from the perturbative one by $(10\div 15)$\%.

\subsection{Computation of the $B_K(\mu)$ parameter}
\label{sec:4f}

The strategy to compute the bag parameter is well known~\cite{Gavela:1988bd} and we only briefly recall it. 
The asymptotic behavior of the two- and three-point correlation function is given by the 
following expressions:
\bea
\label{asym2}
&&\hspace*{-14mm}{\cal C}_{JJ}^{(2)} (\vec p, ; t) = 
 \sum_{\vec x}  \ e^{ -i \vec p \cdot \vec x }\langle \, J (\vec x, t) 
  J^\dagger (0, 0) \, \rangle  \, \stackrel{t\gg 0}{\longrightarrow}\, 
  {{\cal{Z}}_J \over 2 E_J} e^{- E_J t} \ ,\cr
  &&\cr
&&\hspace*{-14mm}{\cal C}^{(3)} (\vec p, \vec q ; t_{P_1}, t_{P_2}; \mu) = 
 \sum_{\vec x \vec y}  \ e^{ i (\vec p \cdot \vec y -  
 \vec q\cdot \vec x )}\langle \, P_5 (\vec x, t_{P_2}) 
  Q(\vec 0, 0; \mu)  P_5^\dagger (\vec y, t_{P_1}) \, \rangle \cr
  &&\cr
\label{3pts}
&& \stackrel{-t_{P_1}, t_{P_2}\gg 0}{\longrightarrow}\, 
  {\sqrt{\cal{Z}}_P \over 2 E_{P}(\vec q)}e^{- E_P(\vec q) t_{P_1} }\  
  \langle \bar P (\vec q)\vert Q(\mu) \vert P (\vec p)\rangle\  {\sqrt{\cal{Z}}_P 
  \over 2 E_{P}(\vec p)} e^{- E_P(\vec p) t_{P_2} } ,
\eea
where the operator $Q(\mu)$ is given by eq.~(\ref{subtr}), 
$\sqrt{{\cal{Z}}_P}\equiv \langle 0\vert P_5 \vert P\rangle$ ($P_5 = i \bar q \gamma_5 q$), 
$\vec p$, $\vec q$ are momenta given to the interacting pseudoscalar mesons. In all our 
simulations, we fix $t_{P_2}=12$, while $t_{P_1}$ is free. 
The elimination of the exponentials in the three-point function is achieved 
by considering the ratio
\bea
\label{ratio}
R (\vec p, \vec q; \mu) \ =\ { {\cal C}^{(3)} (\vec p, \vec q ; t_{P_1}, t_{P_2}; \mu) 
\over Z_A^2 \ {\cal C}_{PP}^{(2)} (\vec p ; t_{P_2}) \ {\cal C}_{PP}^{(2)} (\vec q ; t_{P_1})}\, .
\eea
We also divided by the axial current renormalization 
constant which we computed non-perturbatively by using the method of 
ref.~\cite{Martinelli:1995ty}:~\footnote{Recent works in which $Z_A$ and other 
bilinear quark operators' renormalization constants are computed by using 
the method of ref.~\cite{Martinelli:1995ty}, can be found in refs.~\cite{Gockeler:1999ye,Gimenez:1998ue,Becirevic:2000rv}. 
In refs.~\cite{Luscher:1997jn,Bhattacharya:2000pn}, $Z_A$ has also been computed by using the hadronic 
Ward identities.  The results of the two methods are consistent.} 
\bea
& & \quad \; {\scriptstyle  \rm (a)}  \hspace*{1.1cm} {\scriptstyle \rm (b1)} 
\hspace*{1.1cm} {\scriptstyle \rm (b2)} \hspace*{1.0cm} 
 {\scriptstyle \rm (c)}\cr
Z_A(g_0^2) &=& \left\{ 0.80(1),\; 0.81(1),\; 0.80(1),\; 0.82(1)  \right\} \ 
\,.
\eea

After a careful inspection of the ratio~(\ref{ratio}) for several 
different values of 
$\vec p$ and $\vec q$ on the lattice (a),   
we select the following non-equivalent combinations:
\bea
\label{momenta}
{}\quad \vec p = (0,0,0) &\& & \vec q = \{ (0,0,0), \ (1,0,0) \}\,,\cr
&&\cr
{}\quad \vec p = (1,0,0) &\& & \vec q = \{ (1,0,0),\ (0,1,0) \}\,, 
\eea
where each component is given in units of $(2\pi/L)$.
For the lattices (b1), (b2) and (c), we consider only the first two combinations. 
This is because the temporal axis of these lattices is rather small ($T=40$) 
to accommodate a clear signal when the larger momenta are given to the `kaons'. 
As usual, whenever possible, we average over equivalent kinematical configurations.

Since the Wilson term in the fermion action breaks chiral symmetry, in the chiral 
expansion of the matrix element~(\ref{def1}), one also allow for a 
free term and the one proportional to the quark mass ({\it i.e.} to $m_K^2$):
\bea
\label{chpt}
\langle \bar K^0 (q) \vert Q(\mu) \vert K^0 (p) \rangle = \alpha\ + \ \beta\   m_K^2 + \gamma \ (p\cdot
q) +
\dots \ 
\eea
where the dots stand for higher order terms in expansion.  
The coefficients $\alpha$ and $\beta$ in eq.~(\ref{chpt}) are expected to be consistent with zero 
if the chiral behaviour of the operator $Q(\mu)$ is correct.

To investigate this point we follow ref.~\cite{Gavela:1988bd} and rewrite eq.~(\ref{chpt}) 
to the first order, in the following form:
\bea
\label{fitform1}
R (\vec p, \vec q; \mu )\ =\ \alpha \ +\ \beta  \ \cdot X \ + \ \gamma \ \cdot Y\quad ,
\eea
where the suitable choice for the fit with lattice data is
\bea
\label{XY}
 X\ =\ {8 \over 3}\  \left| \frac{  \displaystyle{\sum_{\vec x} \langle A_0(x) P_5(0)\rangle } }{ 
  \displaystyle{\sum_{\vec x} \langle
P_5(x) P_5(0)\rangle } }\right|^2 \quad \quad
{\rm and} \quad \quad Y\ =\ {p\cdot q\over m_K^2}\ X \quad ,
\eea
It is now clear that the division by $Z_A^2$ in the ratio~(\ref{ratio}) provides the
renormalization
of the axial current ($A_0 = \bar q \gamma_0 \gamma_5 q$) in the quantity $X$, which allows then 
to identify the term proportional to $Y$ as $f_K^2 (p\cdot q)$, where $f_K$ is the (renormalized)
kaon decay constant defined from 
$\langle 0\vert A_\mu\vert K^0(p)\rangle = i f_K p_\mu$.
\begin{table}[h!!] 
\begin{center} 
\begin{tabular}{|c|c|c|c|c|}
\multicolumn{5}{c}{\sf (a)} \\
\hline
{{\phantom{\Huge{l}}}\raisebox{-.1cm}{\phantom{\Huge{j}}}}
\hspace*{-1cm}$\quad \vec p\ ,\ \vec q $ & 
\multicolumn{2}{c|}{\underline{$(0,0,0), (0,0,0)$}} & 
\multicolumn{2}{c|}{\underline{$(0,0,0), (1,0,0)$}} 
\\  \hline 
{{\phantom{\Huge{l}}}\raisebox{-.1cm}{\phantom{\Huge{j}}}}
\hspace*{-6mm} $\kappa_i$ & $Y\equiv X$ & $R(\vec p,\vec q)$& $Y$ & $R(\vec p,\vec q)$ \\ \hline
{\phantom{\Huge{l}}}\raisebox{-.1cm}{\phantom{\Huge{j}}}
\hspace*{-6mm} $ \kappa_1$& 0.068(7)  & 0.040(6) & 0.111(11) & 0.068(10) \\
{\phantom{\Huge{l}}}\raisebox{-.1cm}{\phantom{\Huge{j}}}
\hspace*{-6mm}$\kappa_2$& 0.102(8)  & 0.068(8)& 0.149(12) & 0.102(10)   \\
{\phantom{\Huge{l}}}\raisebox{-.1cm}{\phantom{\Huge{j}}}
\hspace*{-6mm}$\kappa_3$& 0.166(10) & 0.119(12) & 0.217(13) & 0.159(12)   \\  \hline 
\multicolumn{5}{c}{$\hfill $} \\
\hline 
{{\phantom{\Huge{l}}}\raisebox{-.1cm}{\phantom{\Huge{j}}}}
\hspace*{-6mm}$ \vec p\ ,\ \vec q $ & 
\multicolumn{2}{c|}{\underline{$(1,0,0), (0,1,0)$}} & 
\multicolumn{2}{c|}{\underline{$(1,0,0), (1,0,0)$}} 
\\  \hline 
{{\phantom{\Huge{l}}}\raisebox{-.1cm}{\phantom{\Huge{j}}}}
\hspace*{-6mm}$\kappa_i$ & $Y$ & $R(\vec p,\vec q)$& $Y$ & $R(\vec p,\vec q)$ \\ \hline
{\phantom{\Huge{l}}}\raisebox{-.1cm}{\phantom{\Huge{j}}}
\hspace*{-6mm}$\kappa_1$&  0.182(17) & 0.111(17)& 0.066(7) & 0.033(17)  \\
{\phantom{\Huge{l}}}\raisebox{-.1cm}{\phantom{\Huge{j}}}
\hspace*{-6mm}$\kappa_2$& 0.216(17) & 0.150(18)& 0.100(8) & 0.063(16)   \\
{\phantom{\Huge{l}}}\raisebox{-.1cm}{\phantom{\Huge{j}}}
\hspace*{-6mm}$\kappa_3$& 0.283(17) & 0.205(18)& 0.163(10) & 0.108(14)  \\  \hline
\end{tabular}
\caption{\label{tab4}
{\sl \small Numerical values of the quantity $Y$ and of the ratio $R(\vec p,\vec q;\mu)$, computed 
in the RI-MOM scheme at $\mu a= 0.7$ ($\mu = 1.9\ \gev$) for the lattice (a). 
The results are presented for each value of the hopping parameter 
$\kappa_i$ specified in tab.~\ref{tab2}, 
and for four combinations of momenta $\vec p$ and $\vec q$.}}
\end{center} 
\end{table}

Before discussing the physical results, we give in tab.~\ref{tab4} and \ref{tab5} the explicit
values of the quantity $Y$ and the ratio $R(\vec p, \vec q; \mu )$.
We stress that the renormalization and subtraction constants used in this computation are those 
obtained for each value of $\kappa_q$ separately. We have also computed $R(\vec p, \vec q; \mu )$ 
by using the  renormalization and subtraction constants extrapolated to the chiral limit. Since the
extrapolation of these constants to the chiral limit is very smooth,~\footnote{The corresponding plots
can be obtained from the authors.} the net effect on our results
is negligibly small (less than $1\%$). In computing $(p\cdot q)$ we used the 
latticized energy-momentum relation 
\bea
{\rm sinh}^2\left( {E_P(\vec p)\over 2} \right) = {\rm sinh}^2\left( {M_P\over 2
}
\right) + {\rm sin}^2\left( {\vec p\over 2} \right)\ ,
\eea
which describes well our data~\cite{Becirevic:1998jp}.

\begin{table}[h!!] 
\vspace*{.6cm}
\begin{center} 
\begin{tabular}{|c|c|c|c|c|}
\multicolumn{5}{c}{\sf (b1)} \\
\hline 
{{\phantom{\Huge{l}}}\raisebox{-.1cm}{\phantom{\Huge{j}}}}
\hspace*{-1cm}$\quad \vec p\ ,\ \vec q $ & 
\multicolumn{2}{c|}{\underline{$(0,0,0), (0,0,0)$}} & 
\multicolumn{2}{c|}{\underline{$(0,0,0), (1,0,0)$}} 
\\  \hline 
{{\phantom{\Huge{l}}}\raisebox{-.1cm}{\phantom{\Huge{j}}}}
\hspace*{-6mm} $\kappa_i$ & $Y\equiv X$ & $R(\vec p,\vec q)$& $Y$ & $R(\vec p,\vec q)$ \\ \hline
{\phantom{\Huge{l}}}\raisebox{-.1cm}{\phantom{\Huge{j}}}
\hspace*{-6mm} $ \kappa_1$& 0.078(3)  & 0.070(26) & 0.117(6) & 0.101(34) \\
{\phantom{\Huge{l}}}\raisebox{-.1cm}{\phantom{\Huge{j}}}
\hspace*{-6mm}$\kappa_2$& 0.108(3)  & 0.091(27)& 0.151(5) & 0.120(34)   \\
{\phantom{\Huge{l}}}\raisebox{-.1cm}{\phantom{\Huge{j}}}
\hspace*{-6mm}$\kappa_3$& 0.113(3) & 0.113(28)& 0.183(5) & 0.143(30)   \\
{\phantom{\Huge{l}}}\raisebox{-.1cm}{\phantom{\Huge{j}}}
\hspace*{-6mm}$\kappa_4$& 0.200(4) & 0.162(27) & 0.244(5) & 0.189(33)   \\  \hline 
\multicolumn{5}{c}{ } \\
\multicolumn{5}{c}{\sf (b2)} \\
\hline 
{{\phantom{\Huge{l}}}\raisebox{-.1cm}{\phantom{\Huge{j}}}}
\hspace*{-6mm}$ \vec p\ ,\ \vec q $ & 
\multicolumn{2}{c|}{\underline{$(0,0,0), (0,0,0)$}} & 
\multicolumn{2}{c|}{\underline{$(0,0,0), (1,0,0)$}} 
\\  \hline 
{{\phantom{\Huge{l}}}\raisebox{-.1cm}{\phantom{\Huge{j}}}}
\hspace*{-6mm}$\kappa_i$ & $Y\equiv X$ & $R(\vec p,\vec q)$& $Y$ & $R(\vec p,\vec q)$ \\ \hline
{\phantom{\Huge{l}}}\raisebox{-.1cm}{\phantom{\Huge{j}}}
\hspace*{-6mm}$\kappa_1$&  0.060(5) & -0.002(22)& 0.101(10) & 0.013(40)  \\
{\phantom{\Huge{l}}}\raisebox{-.1cm}{\phantom{\Huge{j}}}
\hspace*{-6mm}$\kappa_2$& 0.090(5) & 0.024(22)& 0.134(8) & 0.051(35)   \\
{\phantom{\Huge{l}}}\raisebox{-.1cm}{\phantom{\Huge{j}}}
\hspace*{-6mm}$\kappa_3$& 0.121(5) & 0.051(24)& 0.166(7) & 0.085(34)   \\
{\phantom{\Huge{l}}}\raisebox{-.1cm}{\phantom{\Huge{j}}}
\hspace*{-6mm}$\kappa_4$& 0.183(4) & 0.108(28)& 0.229(5) & 0.148(34)  \\  \hline
\multicolumn{5}{c}{ } \\
\multicolumn{5}{c}{\sf (c)} \\
\hline 
{{\phantom{\Huge{l}}}\raisebox{-.1cm}{\phantom{\Huge{j}}}}
\hspace*{-6mm}$ \vec p\ ,\ \vec q $ & 
\multicolumn{2}{c|}{\underline{$(0,0,0), (0,0,0)$}} & 
\multicolumn{2}{c|}{\underline{$(0,0,0), (1,0,0)$}} 
\\  \hline 
{{\phantom{\Huge{l}}}\raisebox{-.1cm}{\phantom{\Huge{j}}}}
\hspace*{-6mm}$\kappa_i$ & $Y\equiv X$ & $R(\vec p,\vec q)$& $Y$ & $R(\vec p,\vec q)$ \\ \hline
{\phantom{\Huge{l}}}\raisebox{-.1cm}{\phantom{\Huge{j}}}
\hspace*{-6mm}$\kappa_1$&  0.083(14) & 0.056(28)& 0.125(16) & 0.083(27)  \\
{\phantom{\Huge{l}}}\raisebox{-.1cm}{\phantom{\Huge{j}}}
\hspace*{-6mm}$\kappa_2$& 0.100(15) & 0.069(26)& 0.143(16) & 0.097(26)   \\
{\phantom{\Huge{l}}}\raisebox{-.1cm}{\phantom{\Huge{j}}}
\hspace*{-6mm}$\kappa_3$& 0.118(16) & 0.083(24)& 0.163(17) & 0.114(25)   \\
{\phantom{\Huge{l}}}\raisebox{-.1cm}{\phantom{\Huge{j}}}
\hspace*{-6mm}$\kappa_4$& 0.155(17) & 0.113(21)& 0.202(20) & 0.148(24)  \\  \hline
\end{tabular}
\vspace*{.5cm}
\caption{\label{tab5}
{\sl \small The same as in tab.~\ref{tab4} but for lattices (b1), (b2) and (c) in which we consider
two combinations of the momenta $\vec p$ and $\vec q$. }}
\end{center} 
\vspace*{.2cm}
\end{table}
\section{Physical results}

\begin{figure}[h!]
\vspace*{-35mm}
\begin{center}
\begin{tabular}{c c c}
&\epsfxsize16.5cm\epsffile{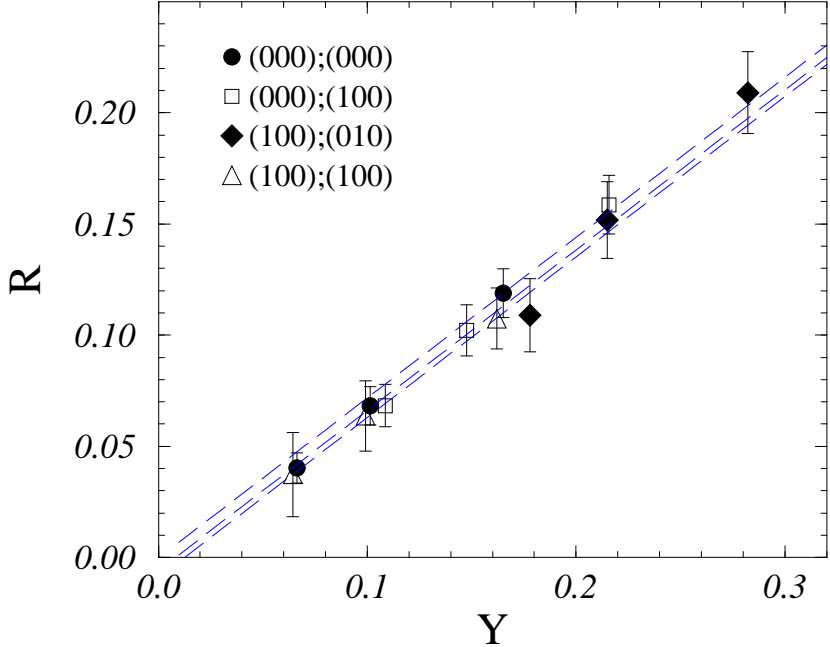} &  \\
\end{tabular}
\vspace*{-15mm}

\caption{\label{fig_fit}{\sl \small Fit of our lattice data (a) to eq.~(\ref{fitform1}).
The shown values of $R(\vec p,\vec q;\mu )$ 
are those corresponding to $\mu a = 0.7$, computed
in the (Landau) RI-MOM scheme. The fitting curves (dashed lines) correspond to three different values of 
$X(\kappa_i)$ in eq.~(\ref{fitform1}).}}
\end{center}
\end{figure}

Our central result is obtained from the fit of our lattice data (a), 
listed in tab.~\ref{tab4}, to the form given in eq.~(\ref{fitform1}). 
This fit is illustrated in fig.~\ref{fig_fit} 
and the results are:
\bea \label{resA}
\mu a = 0.7  &:& \alpha = -0.017(10);\; \beta = 0.10(10);\; B_K^{\rm RI-MOM}(\mu)= 
\gamma = 0.713(67),\cr
&&\cr
\mu a = 1.4 &:& \alpha = -0.009(9);\; \beta =  0.10(10);\; B_K^{\rm RI-MOM}(\mu)=
\gamma = 0.706(65).
\eea 
To convert $B_K^{\rm RI-MOM}(\mu)$ to the RGI form and also to the $\msbar$
(NDR) scheme, in 
which the lattice results are usually presented, we use 
eq.~(\ref{anomalous}). At this point we should decide which coupling to use. 
One choice is to take the value $\alpha_s^{(n_f=0)}(1/a)=0.178$ extracted from the
average plaquette $\langle
P\rangle$, by using the recipe of ref.~\cite{Davies:1997mg}. Another 
(more phenomenological) choice  
is to relate the quenched $B_K^{\rm RI-MOM}(\mu)$ to the physical (unquenched) world by using the 
physical $\alpha_s^{(n_f=4)}(1/a)=0.266$, which is obtained by the NLO evolution 
of $\alpha_s(M_Z)=0.118$. The difference between 
the two conversions is less than $1\%$. In this letter we use 
$\alpha_s^{(n_f=0)}(1/a)$, which leads to
\bea \label{bkA}
B_K^{\msbar}(2\ \gev) = 0.730(68);\quad \hat B_K=1.009(94)\; .
\eea
As mentioned before, the results obtained at two different scales~(\ref{resA}) 
are fully consistent with the
first two coefficient of the anomalous dimension obtained in perturbation theory. 
The numbers given in eq.~(\ref{bkA}) are the average of the
 conversions $B_K^{\rm RI-MOM}(\mu)\to B_K^{\msbar}(2\ \gev)$ from two 
 different values of the scale $\mu$. This is the
 main result of this letter to which we should add the systematic 
 uncertainties which we now discuss.
\noindent
\begin{itemize}
\item[--] Our pseudoscalar mesons are composed of degenerated quarks. The 
quenched $\chi$PT 
suggests that the value of $B_K(\mu)$ may increase by $(3\div 5)\%$ if we 
 work with non-degenerate quarks~\cite{Sharpe:1994dc}. To be on a safe side, we 
 include $+5 \%$ to the systematic uncertainty.
\item[--] We tried to modify the expansion~(\ref{fitform1}), by including the term 
$X \log(m_P^2/f_P^2)$ which is the one that shows up in both the standard and the 
quenched $\chi$PT~\cite{Sharpe:1992ft}. We also tried to modify~(\ref{fitform1})
by adding a term proportional to $X^2$. Any of these two modifications does not 
change the value of  $B_K(\mu)$. 
\item[--]  If instead of the procedure explained above, we use the renormalization 
and subtraction constants extrapolated to the chiral limit before doing the fits 
to eq.~(\ref{fitform1}), the final value for $B_K(\mu)$ gets lower by less than a 
percent. To be conservative, we take $\pm 1\%$ which would also account for the tiny 
difference arising from the choice of $\alpha_s(\mu)$ in the conversion formulae.
\item[--]  Since we work at fixed value of the lattice spacing 
it is difficult to estimate the discretization effects. 
Even though the action we use is free of ${\cal O}(a)$ effects,  
our operator $Q(\mu)$ is unimproved.  To get some idea about the
remaining ${\cal O}(a)$ uncertainty, we repeated the analysis by only improving the 
axial current, $A_0^{(impr.)} = A_0 + c_A \partial_0 P_5$ (with the value of $c_A =
-0.04$~\cite{Luscher:1997ug}). 
This further modifies the final $B_K(\mu)$ by $+1\%$.~\footnote{It can also be
deduced from ref.~\cite{Aoki:1999gw} that the difference between $B_K$ as obtained 
at $\beta \geq 6.1$ and the one extrapolated to the
continuum limit is very small.}

\item[--] Finally, one wonders about the size of the systematic uncertainty that
comes from the use of quenched approximation. 
A comparison between the chiral expansion for the $B_K$ parameter in the quenched 
and in the standard $\chi$PT, indicates that the effect of quenching might be negligible. 
In the next subsection, we will show that our 
unquenched data indeed tend to confirm that expectation.
\end{itemize}
By simply summing up all thee above (small) uncertainties, we end up with:
\bea \label{finalR}
B_K^{\msbar}(2\ \gev) = 0.73(7)^{+0.05}_{-0.01};\quad \hat B_K=1.01(9)^{+0.07}_{-0.01}\; ,
\eea
which is our final result.

\subsection{Unquenching}

\begin{figure}[h!]
\vspace*{-37mm}
\begin{center}
\begin{tabular}{c c c}
&\epsfxsize16.5cm\epsffile{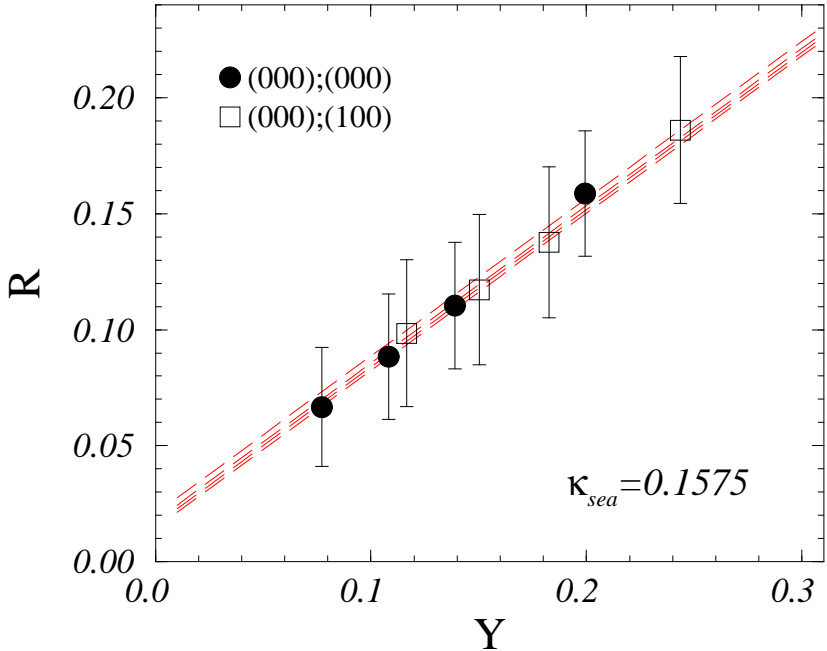} &  \\
\end{tabular}
\vspace*{-15mm}
 
\caption{\label{fig_unq}{\sl \small The same as for
fig.~{\rm \ref{fig_fit}}, but for the partially unquenched data (b1).}}
\end{center}
\end{figure}

We now discuss the results of the analysis of our unquenched data. To our knowledge, 
this is the first attempt to compute the unquenched $B_K(\mu)$, by using  
Wilson fermions. In fact, our results are only partially unquenched (P.U.) which means that 
a valence quark mass is different from the dynamical one. 
Our both sets of data are obtained with $n_f=2$ ((b1) and (b2) in tab.~\ref{tab1}).
As compared to the physical strange quark mass, the dynamical (sea) quark masses
are, $m_{sea}^{(b1)}/m_s \simeq 0.75$, and $m_{sea}^{(b2)}/m_s \simeq 0.50$. 
The results of the fit to eq.~(\ref{fitform1}) are given in tab.~\ref{tab6}
and illustrated in fig.~(\ref{fig_unq}).
\begin{table}[h!]
\begin{center} 
\begin{tabular}{|c|c|c|c|c|}
\hline 
{\phantom{\Huge{l}}}\raisebox{-.1cm}{\phantom{\Huge{j}}}
{\sl Label} & $\quad \mu a \quad $ & $\alpha$ & $\beta$ & $\gamma = B_K^{\rm RI-MOM}(\mu)$\\ \hline \hline 
{\phantom{\Huge{l}}}\raisebox{-.1cm}{\phantom{\Huge{j}}}
\underline{\sf{(b1)}}& $0.7$ & 0.01(3)  & 0.06(19)  & 0.67(20)  \\
{\phantom{\Huge{l}}}\raisebox{-.1cm}{\phantom{\Huge{j}}}
                     & $1.4$ & -0.01(3)  &0.02(16)   & 0.65(17)  \\ \hline 
{\phantom{\Huge{l}}}\raisebox{-.1cm}{\phantom{\Huge{j}}}
\underline{\sf{(b2)}}& $0.7$ & -0.06(3)  & 0.28(24)  & 0.66(26)  \\
{\phantom{\Huge{l}}}\raisebox{-.1cm}{\phantom{\Huge{j}}}
                     & $1.4$ & -0.08(3)  & 0.21(24)  & 0.65(26)   \\ \hline 
{\phantom{\Huge{l}}}\raisebox{-.1cm}{\phantom{\Huge{j}}}
\underline{\sf{(c)}}& $0.7$ & -0.01(3)  & 0.13(19)  &  0.69(18)  \\
{\phantom{\Huge{l}}}\raisebox{-.1cm}{\phantom{\Huge{j}}}
                     & $1.4$ & -0.02(3)  & 0.09(19)  &  0.69(18) \\ \hline 
\end{tabular}
\caption{\label{tab6}
\small{\sl Result of the fit of our data obtained from the computation on the lattices 
(b1), (b2) and (c), to the form given in eq.~\ref{fitform1}).}}
\end{center}
\end{table}
                                                                    As in the previous subsection, we convert the obtained results to the RGI form and to 
the $\msbar$ scheme. We use $\alpha_s^{(n_f=2)}(1/a)= 0.223$ and $0.222$ for lattices (b1) and (b2) 
respectively, and $\alpha_s^{(n_f=0)}(1/a)= 0.187$ for the lattice (c), which as
in the previous subsection are obtained from the average plaquettes.
We finally have:
\bea
{\rm (b1)}&& B_K^{\msbar}(2\ \gev) = 0.67(19);\quad \hat B_K = 0.94(26)\ ,\cr
&& \cr
{\rm (b2)}&& B_K^{\msbar}(2\ \gev) = 0.67(26);\quad \hat B_K = 0.94(37)\ ,\cr
&& \cr
{\rm (c)}&& B_K^{\msbar}(2\ \gev) = 0.70(18);\quad \hat B_K = 0.97(25)\ .
\eea
The statistical errors in these results are large. Systematics due to the lattice geometry
is the same for both sets of data (quenched and unquenched), so that the difference would
indicate the quenching errors.

First of all, from the above numbers ((b1) and (b2)) we do not observe any
difference in $B_K$ as the value of the sea quark mass changes. 
In principle, to get a fully unquenched result, one should work with more than two values of the dynamical quark mass, and then 
extrapolate $B_K$ in such a way that the dynamical quark mass corresponds to the 
physical up/down quark. Since we do not see any change in $B_K(\mu)$ as the value of the sea quark
mass change, we can not distinguish partially from the fully unquenched value for $B_K(\mu)$.

Even more, we do not see any 
significant effect due to the unquenching, {\it i.e.} by switching off the 
quark loops from the simulations
(b1) and (b2), we get the results from the run (c), which totally agrees with the unquenched ones.
This observation is on the same line with the conclusions drawn from the 
comparison of the standard and the (partially) quenched 
$\chi$PT~\cite{Sharpe:1992ft,Golterman:1998st}. 
In ref.~\cite{Kilcup:1997hp}, however, by using the staggered fermions, an enhancement 
of $B_K$ of the order of $(5\pm 2)\%$ due to the unquenching has been reported. 
Obviously, with a poor statistical 
quality of our data, we are unable to confirm this result.~\footnote{In fact, 
the enhancement in $B_K$ due to the unquenching $\sim 5$\%, as claimed by 
the authors of 
ref.~\cite{Kilcup:1997hp},  comes from
their simulations with $n_f=4$, 
while their data with $n_f=2$ almost coincide with the quenched ones.}

In conclusion, we do not add any systematic uncertainty to our 
result~(\ref{finalR}) due to the use of quenched approximation.

\section{Conclusion and prospects}
In this letter, we computed the $B_K$ parameter by using ${\cal O}(a)$ 
improved Wilson action but without improving the corresponding four fermion operator.
From the non-perturbatively renormalized data, we obtained $B_K^{\msbar}(2\ \gev)=0.73(7)^{+0.05}_{-0.01}$,
which confirms the tendency of larger values for this quantity when working with 
the Wilson fermions~\cite{Aoki:1999gw,Conti:1998qk,Gupta:1997yt,Lellouch:1999sg}. 
In order to examine the effect of quenching, we computed $B_K$ with
dynamical fermions ($n_f=2$). Even though the statistical 
quality of our unquenched results is poor, we do not see any significant deviation 
of the parameter $B_K$ as compared to the one obtained in the quenched simulation. 
To quantify the small quenching errors (if any!) it is obviously important 
to increase the statistics of the unquenched data. 

The main systematic uncertainty, which we could not estimate, is expected 
to come from the
computation of the bare operator, {\it i.e.} from the subtractions of the 
effects of mixing with operators of the same dimension (peculiarity of the Wilson 
lattice regularization). 
This can be completely avoided if one follows the proposal of ref.~\cite{Becirevic:2000cy},
namely by using the Ward identity 
to relate $Q(\mu)=VV+AA$ to the
parity violating operator ${\cal Q}(\mu)=VA+AV$ for which no subtraction is needed. 
Alternatively, 
one can use the twisted mass fermions, as proposed in ref.~\cite{Frezzotti:2000vv}. 
It will be very interesting
to see if the results for $B_K$ without subtractions will confirm the 
values obtained in this letter.

\section*{Acknowledgements} We thank T$\chi$L Collaboration for providing us 
with the unquenched gauge field configurations. 
We are grateful to V.~Gim\'enez, V.~Lubicz and G.~Martinelli for many 
discussions, and to Ph.~Boucaud, L.~Giusti and A.~Vladikas for 
their comments on the manuscript.  
D.B. also thanks INFN for the financial support.

\end{document}